\documentclass[a4paper,UKenglish,cleveref, autoref, thm-restate]{lipics-v2019}

\usepackage{tikz}
\usetikzlibrary{calc}
\usetikzlibrary{patterns}
\usepackage{csquotes}
\usepackage{mdframed}

\crefname{observation}{observation}{observations}
\Crefname{observation}{Observation}{Observations}

\title{A Compendium of Reductions: reductions.network}

\titlerunning{reductions.network} 

\author{Christoph Grüne}{Department of Computer Science, RWTH Aachen University, Germany}{gruene@algo.rwth-aachen.de}{https://orcid.org/0000-0002-7789-8870}{Funded by the German Research Foundation (DFG) – GRK 2236/2.}
\author{Femke Pfaue}{Department of Computer Science, RWTH Aachen University, Germany}{femke.pfaue@rwth-aachen.de}{}{Funded by the German Research Foundation (DFG) – GRK 2236/2.}

\authorrunning{C. Grüne and F. Pfaue} 

\Copyright{Christoph Grüne and Femke Pfaue} 



\category{} 

\relatedversion{} 

\supplement{}

\nolinenumbers 

\hideLIPIcs  

\EventEditors{John Q. Open and Joan R. Access}
\EventNoEds{2}
\EventLongTitle{42nd Conference on Very Important Topics (CVIT 2016)}
\EventShortTitle{CVIT 2016}
\EventAcronym{CVIT}
\EventYear{2016}
\EventDate{December 24--27, 2016}
\EventLocation{Little Whinging, United Kingdom}
\EventLogo{}
\SeriesVolume{42}
\ArticleNo{23}

\definecolor{darkgreen}{RGB}{0,128,0}
\definecolor{darkred}{RGB}{128,0,0}

\newcommand{\NP}{\textsf{NP}}

\newcommand{\NPS}{\textsf{SSP-NP}}
\newcommand{\sharpP}{\textsf{\#P}}
\newcommand{\Wone}{\textsf{W[1]}}
\newcommand{\Wtwo}{\textsf{W[2]}}

\newcommand{\rednet}{\textit{reductions.network}}

\begin{document}

\maketitle
\begin{abstract}
    The website $\rednet$ serves as a comprehensive database for exploring problems and reductions between them.
It presents several complexity classes in the form of an interconnected graph where problems are represented as vertices, while edges represent reductions between them.
This graphical perspective allows for identifying problem clusters and simplifying finding problem candidates to reduce from.
Moreover, users can easily search for existing problems via a dedicated search bar, and various filters allow them to focus on specific subgraphs of interest.
The design of the website enables users to contribute by adding new problems and reductions to the database.
Furthermore, the software architecture allows for the integration of additional graphs corresponding to new complexity classes.
In the current state, the following networks with their respective complexity classes are included:
\begin{itemize}
    \item \textbf{classical complexity} including the classes $\NP$, $\sharpP$, and $\NPS$
    \item \textbf{parameterized complexity} including the classes $\Wone, \Wtwo$
    \item \textbf{gap-preserving reductions} under the PCP-Theorem and the Unique Games Conjecture.
\end{itemize}
    
\end{abstract}

\newpage
\section{Introduction}

In computational complexity, we are interested in grouping problems into complexity classes in order to understand the resource consumption (time, space, etc.) to solve such problems.
In this paper, we present the website $\rednet$, which is a comprehensive resource designed to serve as a central repository for information on various computational problems and reductions between them.
Drawing inspiration from the seminal work of Garey and Johnson \cite{DBLP:books/fm/GareyJ79} and the website dedicated to NP Optimization and Approximation \cite{websiteKann}, this resource places a particular emphasis on the concept of reductions.

The core feature of this website are graph networks that allow users to explore problems and their connections induced by corresponding reductions.
Each vertex in such a graph represents a distinct problem.
By clicking on a vertex, users can access additional information on the corresponding problem, such as a formal description and references.
All problems in one network are searchable through a search bar.
Furthermore, edges connecting the vertices represent reductions between the corresponding problems.
Users can also access further information by clicking on the edges, such as a description of the transformation and further references.
Reductions may also include graphical examples of small instances, such that they can be easily understood.
In order to show connections between reductions that additionally adhere to certain properties, the users are able to filter problems and reductions for said properties, allowing them to focus on specific sub-networks.
An example are parsimonious reductions that are a subclass of typical polynomial-time reductions in the realm of the classical classes $\NP$ and $\sharpP$.

Through this network lens, it is possible to identify problem clusters that share similar reduction characteristics.
This helps researchers in finding suitable problems from which they can reduce to others, and it provides easy access to information regarding similar reductions across similar problems.
Moreover, it is a valuable didactical tool, providing users with essential resources and interactive features that enhance their understanding of computational problems and reductions.

The website is designed to be extendable.
Users are invited to contribute by adding new problems and reductions, enhancing existing entries with additional information, or even creating entirely new networks.
Further information on how to contribute is presented in \Cref{sec:contribution}.
These user-generated additions are stored in a data repository \cite{reductioncompendium-data}, which synchronizes with our website's database.

The development of this website builds upon the bachelor's thesis of Pfaue \cite{DBLP:journals/corr/abs-2411-05796} and currently includes results from the paper by Grüne and Wulf \cite{DBLP:conf/ipco/GruneW25} and the theses by Bartlett \cite{DBLP:journals/corr/abs-2506-12255}, Faour \cite{faour}, Verma \cite{verma}, and He \cite{he}.
By presenting these insights, we aim to create an invaluable tool for researchers engaged in exploring computational problems and reductions between them.

\subsection{Related Work}
In the last decades, different compendia for problems or in general complexity classes were developed.
The most important is the compendium of $\NP$ problems by Garey and Johnson \cite{DBLP:books/fm/GareyJ79} from 1979, which is widely known.
In the 90s, an $\NP$ optimization and approximation compendium was introduced in the book \cite{DBLP:books/lib/Ausiello99}, which is also presented on the web \cite{DBLP:conf/random/CrescenziK97,websiteKann}.
Moreover, Greenlaw, Hoover, and Ruzzo's book \cite{greenlaw1995limits} provides a compendium for $\sf P$-complete problems, and Umans and Schaeffer built up a compendium on problems in the lower levels of the polynomial hierarchy.

In the realm of parameterized complexity, Downey and Fellows' books \cite{DBLP:series/mcs/DowneyF99,DBLP:series/txcs/DowneyF13} include a compendium for problems in the complexity classes of the $\textsf{W}$-hierarchy and further parameterized complexity classes.
Loosely connect to this project are the complexity zoo \cite{aaronson2005complexity} (a collection of complexity classes), the house of graphs \cite{coolsaet2023house} (a searchable database for interesting graphs), the arcane algorithm archive \cite{arcaneAlgorithm} (a collection of important algorithms), the website about graph classes \cite{graphClasses} (a collection of complexity results for problems on certain graph classes), and the HOPS website \cite{hops} (a collection of problem parameters and their hierarchy).

\section{Functionality}

\begin{figure}[!ht]
    \centering
    \includegraphics[width=0.67\linewidth]{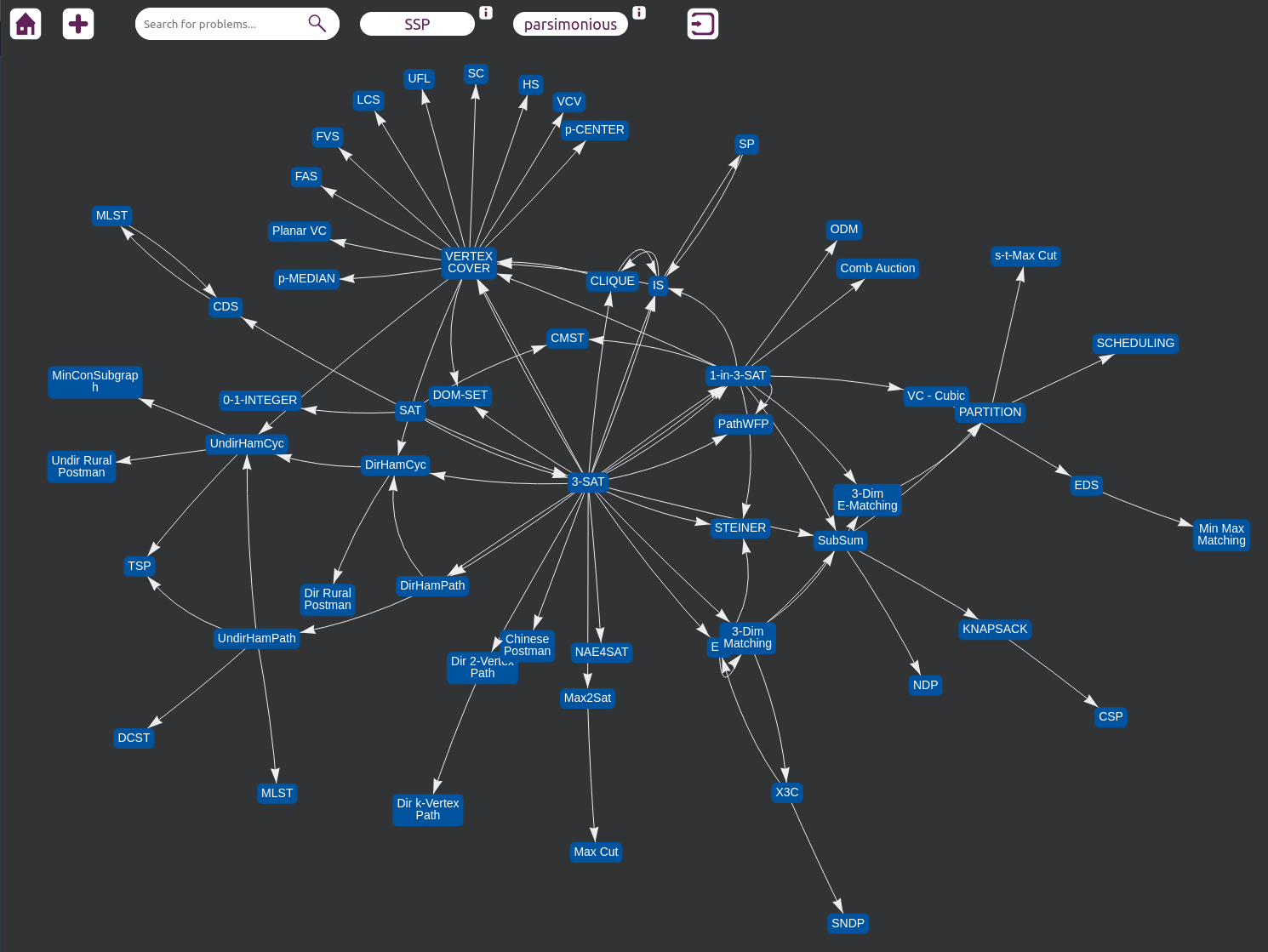}
    \caption{The classical network.}
    \label{fig:classical-network}
\end{figure}

The website displays several types of problems and reductions in graphs, in which the vertices represent the problems and the edges represent the reductions between the problems.
The vertices are labeled with the abbreviation of the problem.
In \Cref{fig:classical-network}, the network of classical problems is displayed.

\begin{figure}[!ht]
    \centering
    \includegraphics[width=0.6\linewidth]{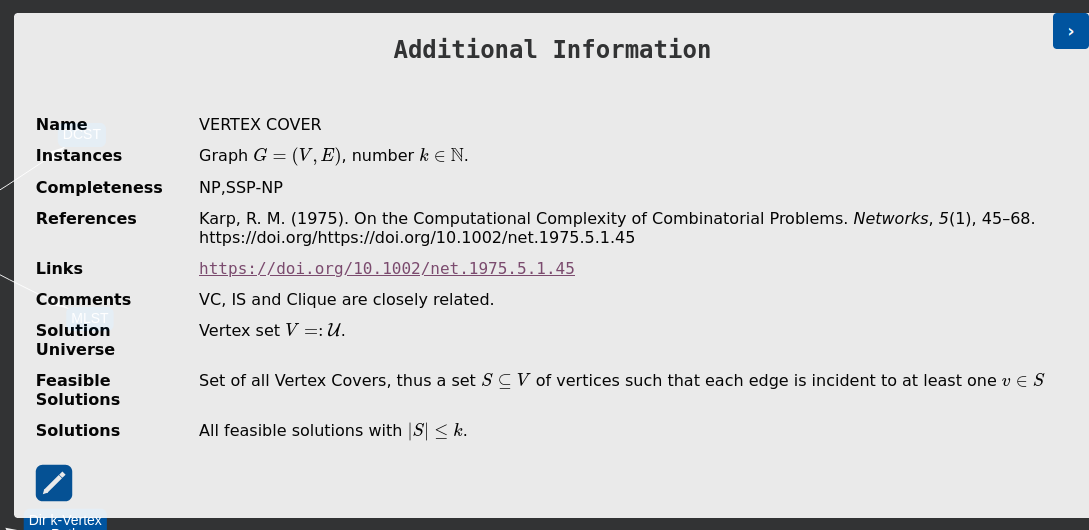}
    \caption{The Window for the problem \textsc{Vertex Cover} in the classical network.}
    \label{fig:classical-network:vertex-cover}
\end{figure}

Clicking on a vertex in the graph opens up a window displaying further information on the problem and additionally highlights the vertex and the incident edges.
An example for the problem \textsc{Vertex Cover} can be found in \Cref{fig:classical-network:vertex-cover}.

\begin{figure}[!ht]
    \centering
    \includegraphics[width=0.6\linewidth]{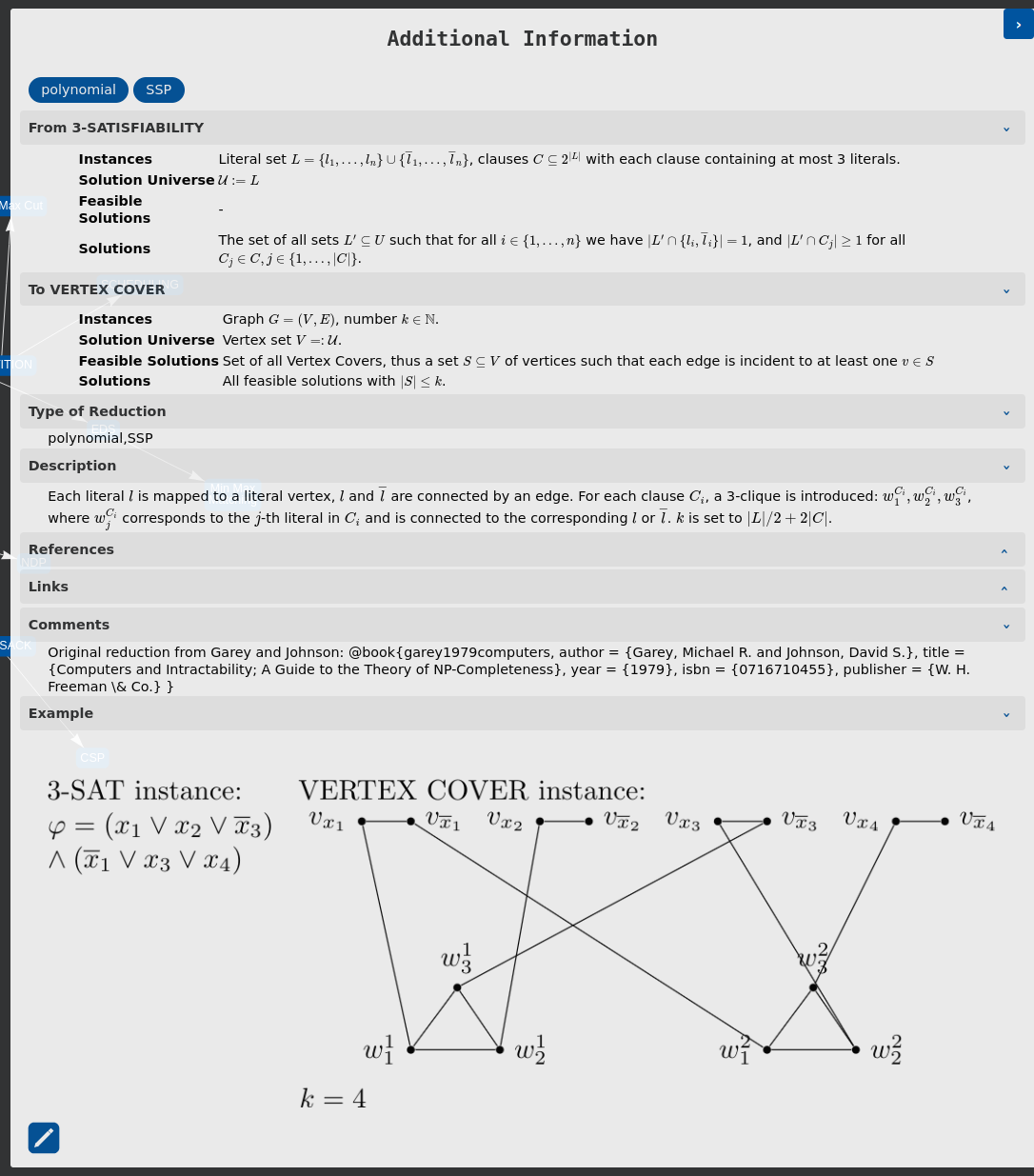}
    \caption{The window for the reduction from \textsc{3Satisfiability} to \textsc{Vertex Cover} in the classical network.}
    \label{fig:classical-network:reduction:sat-vertex-cover}
\end{figure}

It is also possible to click on an edge to open up a window displaying further information on the corresponding reduction, as well as a summary of the corresponding problems.
An example of the reduction from \textsc{Satisfiability} to \textsc{Vertex Cover} can be found in \Cref{fig:classical-network:reduction:sat-vertex-cover}.

\begin{figure}[!ht]
    \centering
    \includegraphics[width=0.67\linewidth]{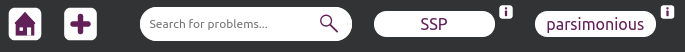}
    \caption{The navigation toolbar for the classical network.}
    \label{fig:navigation-tools}
\end{figure}

Each network is searchable and navigable with the following tools, which are displayed in \Cref{fig:navigation-tools}.
First, there is a search bar that can be utilized to efficiently locate problems within a network.
Users have the capability to search for a specific problem using both its main name and any alternative names that may apply.
Once the search is initiated, the network is focused on the problem.
Additionally, the information window concerning the problem opens up.

Second, a filter function for selected properties is implemented to show sub-networks within a network.
Problems are grouped based on their completeness, while reductions are categorized according to their specific properties. This enables users to filter both problems and reductions effectively, streamlining the search process.
The toolbar features clickable buttons that allow users to activate filters corresponding to particular properties.
For instance, when selecting the "parsimonious" property, only parsimonious reductions will be displayed alongside the related problems that have a counting version classified as \sharpP-complete.

\subsection{The Networks}
Currently, the following networks are implemented in the website.
\begin{itemize}
    \item classical problems in the realm of $\NP$ \cite{DBLP:books/fm/GareyJ79} and $\sharpP$ \cite{DBLP:journals/tcs/Valiant79} and further extensions such as $\NPS$ \cite{DBLP:conf/ipco/GruneW25} 
    \item parameterized problems classes: W-hierarchy \cite{DBLP:series/txcs/DowneyF13}
    \item approximation problems for which gap-preserving reductions from the PCP-Theorem \cite{DBLP:journals/jacm/AroraLMSS98} or the Unique Games Conjecture \cite{DBLP:conf/stoc/Khot02a} exist.
\end{itemize}

\section{Software and Tools}
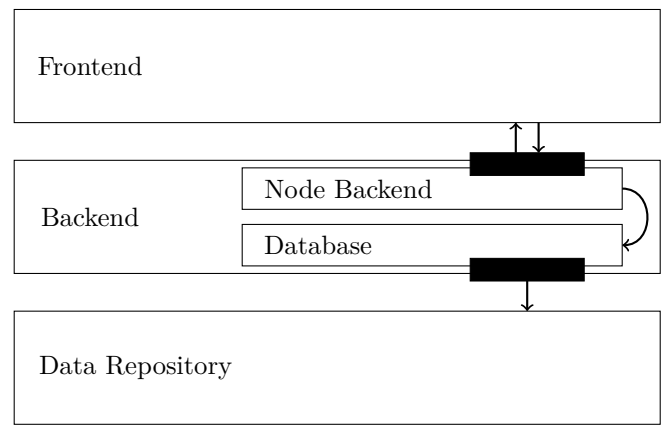
\begin{figure}[!ht]
    \centering
    \scalebox{1}{
        \begin{tikzpicture}
            \draw (0,0) rectangle (8.5,1.5);
            \node at (1.6,0.75) {Data Repository};

            \draw (0,2) rectangle (8.5,3.5);
            \node at (1,2.75) {Backend};
            
            \draw (3,2.1) rectangle (8,2.65);
            \node at (4,2.375) {Database};
            \draw (3,2.85) rectangle (8,3.4);
            \node at (4.4,3.125) {Node Backend};
            \draw[thick,looseness=1.5,out=0,in=0] (8,2.375) edge[<-] (8,3.125);
    
            \draw[fill=black] (6,1.9) rectangle (7.5,2.2);
            \draw[<-,thick] (6.75,1.5) -- (6.75,1.9);
            \draw[fill=black] (6,3.3) rectangle (7.5,3.6);
            \draw[->,thick] (6.6,3.6) -- (6.6,4);
            \draw[->,thick] (6.9,4) -- (6.9,3.6);
            
            \draw (0,4) rectangle (8.5,5.5);
            \node at (1,4.75) {Frontend};
        \end{tikzpicture}
    }
    \caption{The layered architecture of the software.}
    \label{fig:software-architecture}
\end{figure}

The whole software architecture is split into three parts, see also \Cref{fig:software-architecture}.
First, there is a data repository, in which all data on problems and reductions are stored.
This repository is publicly accessible on the RWTH GitLab Server \cite{reductioncompendium-code}, thus enabling users to contribute to the website by adding new information to existing problems and reductions or adding new problems and reductions.
Second, the website backend serves the data to the client over an API.
Accordingly, the data repository is synchronized with the backend to provide information to the frontend.
Third, the frontend calls the API to display the networks and the information on problems and reductions to the user.

The website is built using Node.js \cite{node-js}, a runtime environment for JavaScript that enables the creation of servers and applications.
Node.js allows JavaScript code to run outside of a browser environment, employing an asynchronous design that enhances efficiency for I/O-intensive tasks.
This approach ensures that the thread remains unblocked while awaiting the return value of executed I/O operations.
One significant advantage of using Node.js is the ability to utilize JavaScript on both the backend and frontend.
This consistency in programming language improves maintenance and readability throughout the project.
Additionally, Node.js offers a wide array of libraries and functionalities, either built-in or accessible through npm (Node Package Manager).
Thus, it helps to avoid redundancy by allowing developers to use features already implemented by others.
The project has been developed without employing a client-side framework such as React or Angular.
Given its relatively simple structure, consisting of only a few pages, a framework was deemed unnecessary for this particular application.

\subsection{The Data Repository}
The data for the project is stored in a Git repository \cite{reductioncompendium-data}, where the data for each problem and reduction is stored in a separate file.
These files are encoded in Markdown format, with a specific structure that includes sections for the properties of problems and reductions.
In these Markdown files, GitLab supports inline TeX math using the \$ and \$\$ symbols, providing a reasonable approximation for the website's output.

The structure of each file is clearly defined: fields are indicated with \# as headings, while the corresponding content follows directly beneath each heading, for example
\begin{quote}
    \# name\\
    Vertex Cover
\end{quote}
Users have the flexibility to enter new data into existing files or add entirely new files to the appropriate network by placing them in designated folders.
The naming conventions for these files are also important to maintain organization:
On the top level, the folders for each implemented network can be found; specifically, approximation, classic, and parameterized.
In each of the network folders, the folders for problems and reductions can be found, in which the actual Markdown files are located.

To ensure consistency and adherence to the defined structure, continuous integration (CI) through Git is utilized.
This system automatically checks whether the Markdown files comply with the specified format, allowing users to receive immediate feedback if any issues arise during their submissions.
After a pull request containing new data is accepted and merged, the data is synchronized with the live database in the backend of the website in regular time intervals such that all updates are provided to users accessing the site.

\subsection{Backend}
The backend of the website is powered by a MariaDB \cite{mariadb} database, which serves requests specified through an API.
This open-source relational database management system utilizes SQL to store all information related to problems and reductions.
In this project, the mariadb package is employed to establish a connection with the database and execute queries efficiently.
It allows for the creation of a pool of connections, enabling multiple queries to run in parallel without the need for initiating new connections each time.

To facilitate web interactions, Express \cite{express-js} — a web framework for Node.js — is utilized.
Express manages routes, creates APIs, and offers various middleware modules essential for the project's functionality.
Specifically, it serves files, manages user sessions with express-session, validates incoming data using express-validator, limits access to prevent flooding through express-rate-limit, and handles routing and API creation.

The API is designed to manage requests related to complete networks as well as specific information on individual problems or reductions, including filtering capabilities.
Additionally, data from the repository is regularly pulled to synchronize with the database, ensuring that all information remains up-to-date and accurate for users accessing the website.

\subsection{Frontend}
The frontend of the project is developed using the Node.js framework \cite{node-js}, which incorporates HTML, CSS, JavaScript, and EJS.
The primary focus of the frontend is to present data visually in the form of graphs, which are generated using the Vis.js library \cite{vis-js,vis-js-github}.
Vis.js stands out as the most crucial library utilized in this project.
This browser-based visualization library is able to handle large volumes of dynamic data, making it particularly suitable for creating a compendium website that may contain thousands of entries.
Originally developed by Almende B.V. \cite{almende}, Vis.js is now maintained by a community of contributors.
In this project, the network library component of Vis.js is employed to produce the network visualizations on the website.
Several key factors influenced the choice of Vis.js as the visualization library:
(1) Scalability - it can efficiently manage large datasets -
(2) Manipulation - it allows for dynamic changes to both data and appearance -
(3) Interaction - users can click on elements to access more detailed information -
(4) Open Source - it is free to use.

Vis.js meets all these criteria and benefits from extensive documentation, support, and examples due to its widespread use.
Additionally, being a JavaScript library facilitates easy integration into the website, avoiding complications associated with alternatives based on other programming languages - such as Pyvis \cite{pyvis} (a Python library) or NetworkD3 (which has roots in R).

The features provided by Vis.js include physics simulation for creating dynamic networks and capabilities for manipulating edges and nodes in real time, such as changing colors or adding and removing nodes and edges.
The physics simulation helps to approximate clusters within the network and allows the user to drag and reposition vertices and edges of the graph.
Key functions implemented in this project include initializing the network with specific options and data, retrieving nodes and edges, updating and removing elements without requiring a full reload, and centering the view on specific nodes using the focus method, which is essential for implementing search functionalities.
Moreover, Vis.js offers useful features like click listeners for nodes and edges.
By clicking on any vertex or edge, users can access a box displaying all relevant information retrieved from the database via an API.

For managing citations effectively within the frontend, Citation-js \cite{citation-js} is employed to render citations in various formats accurately using BibTeX.
While it has primarily been tested with BibTeX format, it should also accommodate other standard citation formats such as BibJSON or BibLaTeX without issues.
Mathematical formulas are rendered using MathJax \cite{mathjax} in LaTeX format to ensure correct display across devices; importantly, MathJax is accessible to screen readers.
Although it may not be the fastest rendering solution due to running on the frontend, its performance was satisfactory during development since only small amounts of LaTeX code were needed at any given time, resulting in no noticeable delays or performance problems while also reducing server workload.

\section{How to Contribute?}
\label{sec:contribution}
To contribute to the project, please navigate to the GitLab data repository \cite{reductioncompendium-data}.
The README file of the repository explains how data can be added.
The repository is synchronized each time a pull request is merged, ensuring that all contributions are up-to-date.
It is essential that all contributions adhere to the specified format, which is checked by the continuous integration (CI) and gives corresponding feedback.
However, if you encounter any unexpected issues, please post an issue on the Git repository for assistance.

The software architecture of the website is designed to facilitate the easy addition of new networks. To do so, a new data structure must be created for the new problems and reductions. This process requires several key steps:
\begin{description}
    \item[Data Repository] A new directory must be established within the GitLab data repository, along with an extension of the GitLab CI to accommodate these changes.
    \item[Backend] Additional tables need to be added to the database, along with new database queries that correspond to the newly defined structures.
    Furthermore, it will be necessary to extend the API to handle requests related to these additions.
    \item[Frontend] Finally, a corresponding link must be added on the homepage that directs users to the new network webpage.
\end{description}
If you want to contribute a new network, please open up an issue in the GitLab code repository \cite{reductioncompendium-code}.
We will assist with any possible additions.

\newpage

\bibliography{bibliography,websites}

\newpage

\end{document}